\documentclass[prb,twocolumn,showpacs,preprintnumbers,amsmath,amssymb,superscriptadress,bibnotes,superscriptaddress]{{revtex4-2}}% Physical Review B

\usepackage{color}
\usepackage{ulem}
\usepackage[pdftex]{graphicx}
\usepackage{dcolumn}% Align table columns on decimal point
\usepackage{bm}% bold math

\usepackage{array}
\usepackage{float}
\usepackage{units}
\usepackage{multirow}
\usepackage{amsmath}
\usepackage{stmaryrd}
\usepackage{setspace}
\usepackage{here}

\usepackage{mathrsfs}

\usepackage{natbib}

\begin{document}

%\preprint{APS/123-QED}

\title{Fermi-liquid state in $T$*-type La$_{1-x/2}$Eu$_{1-x/2}$Sr$_x$CuO$_4$ revealed via element substitution effects on magnetism}% Force line breaks with \\

\author{Takanori Taniguchi}
\email{taka.taniguchi@imr.tohoku.ac.jp}
\affiliation{Institute for Materials Research, Tohoku University, Katahira, Sendai 980-8577, Japan}
\author{Kota Kudo}
\author{Shun Asano}
\author{Motofumi Takahama}
\affiliation{Department of Physics, Tohoku University, Aoba, Sendai 980-8578, Japan}
\affiliation{Institute for Materials Research, Tohoku University, Katahira, Sendai 980-8577, Japan}
\author{Isao Watanabe}
\altaffiliation{Present address: Nuclear Structure Research Group, RIKEN Nishina Center for Accelerator-Based Science, RIKEN, Wako, Saitama 351-0198, Japan}
\affiliation{Meson Science Laboratory, RIKEN Nishina Center for Accelerator-Based Science, RIKEN, Wako, Saitama 351-0198, Japan}
\author{Akihiro Koda}
\affiliation{Institute of Materials Structure Science, High Energy Accelerator Research Organization, Tsukuba, Ibaraki 305-0801, Japan}
\author{Masaki Fujita}
\email{fujita@imr.tohoku.ac.jp}
\affiliation{Institute for Materials Research, Tohoku University, Katahira, Sendai 980-8577, Japan}

\date{\today}% It is always \today, today,

\begin{abstract}
Despite its unique structural features, the magnetism of single-layered cuprate with five oxygen coordination ($T$*-type structure) has not been investigated thus far. 
Here, we report the results of muon spin relaxation and magnetic susceptibility measurements to elucidate the magnetism of $T$*-type La$_{1-x/2}$Eu$_{1-x/2}$Sr$_x$CuO$_4$ (LESCO) via magnetic Fe- and non-magnetic Zn-substitution. 
We clarified the inducement of the spin-glass (SG)-like magnetically ordered state in La$_{1-x/2}$Eu$_{1-x/2}$Sr$_x$Cu$_y$Fe$_{1-y}$O$_4$ with $x = 0.24 + y$, and the non-magnetic state in La$_{1-x/2}$Eu$_{1-x/2}$Sr$_x$Cu$_y$Zn$_{1-y}$O$_4$ with $x$ = 0.24 after the suppression of superconductivity for $y$ $\geq$ 0.025. The SG state lies below $\sim$7 K in a wide Sr concentration range between 0.19 and 0.34 in 5$\%$ Fe-substituted LESCO. 
The short-range SG state is consistent with that originating from the Ruderman-Kittel-Kasuya-Yosida interaction in a metallic state. Thus, the results provide the first evidence for Fermi liquid (FL) state in the pristine $T$*-type LESCO. Taking into account the results of an oxygen $K$-edge X-ray absorption spectroscopy measurement $[$J. Phys. Soc. Jpn. {\bf 89}, 075002 (2020)$]$ reporting the actual hole concentrations in LESCO, our results demonstrate the existence of the FL state in a lower hole-concentration region, compared to that in $T$-type La$_{2-x}$Sr$_x$CuO$_4$. The emergence of the FL state in a lower hole-concentration region is possibly associated with a smaller charge transfer gap energy in the parent material with five oxygen coordination. 
\end{abstract}


\maketitle

\section{\label{sec:level1}Introduction}
A recent study on superconductivity in the $RE_2$CuO$_4$ ($RE$ = rare earth) family has reported a variety of ground states due to different oxygen coordination around Cu$^{2+}$~\cite{Krockenberger2014, Adachi2017}. It is well known that $T$-type La$_2$CuO$_4$ with six coordination is a Mott insulator that exhibits superconductivity with hole doping. $T^{\prime}$-type $RE_2$CuO$_4$ with four planar coordination has also been considered to be the parent Mott insulator of electron-doped superconductors~\cite{Tokura1990, Arima1991, Uchida1991, Onose2004}. However, Yamamoto et al. reported the emergence of superconductivity in a thin film of $T^{\prime}$-type $RE_2$CuO$_4$~\cite{Tsukada2005, Matsumoto2009a, Matsumoto2009b}, leading to the study of the actual ground state in $RE_2$CuO$_4$ with different coordinations~\cite{Asai2011, Takamatsu2012}. 
A theoretical study demonstrated a decrease in the electron correlation strength with increasing distance between the apical oxygen and the CuO$_2$ plane, supporting the metallic nature in $T^{\prime}$-type cuprates~\cite{Jang2016}. In real materials, partially existing apical oxygen atoms in the as-sintered (AS) $T^{\prime}$-type compounds are considered to inhibit superconductivity, owing to the introduction of a random electronic potential on the CuO$_2$ plane~\cite{Radaelli1994c, Xu1996, Schultz1996}. Thus, a reduction annealing procedure is necessary for the complete removal of excess oxygen to induce superconductivity~\cite{Krockenberger2013}. 

In another structural isomer of $RE_2$CuO$_4$ with five coordination ($T$*-type structure), superconductivity appears with hole doping and post annealing~\cite{Akimitsu1988, Sawa1989}. $T$*-type cuprates have an intermediate crystal structure between the $T$ and $T^{\prime}$-type cuprates, which contains a CuO$_5$ pyramid in the unit cell. Although studies on the physical properties of $T$*-type cuprates is limited, Kojima et al. reported weak magnetism in $T$*-type La$_2$CuO$_4$~\cite{Kojima2014}, which is qualitatively similar to that in superconducting (SC) $T^{\prime}$-type $RE_2$CuO$_4$~\cite{Kojima2014, Adachi2013, Kawamata2018}. Thus, T*-type cuprates are a notable system for studying the relationship between physical properties and oxygen coordination. However, owing to the limited availability of high-quality samples~\cite{Cheong1989, Fisk1989}, $T$*-type cuprates have not been studied thoroughly. 

Recently, we reported a magnetic and SC phase diagram of $T$*-type La$_{1-x/2}$Eu$_{1-x/2}$Sr$_x$CuO$_4$ (LESCO) with $x$ ranging from 0.14 to 0.26 by muon spin relaxation ($\mu$SR), magnetic susceptibility, and electrical resistivity measurements~\cite{Fujita2018, Asano2019}. All the AS and oxidation annealed (OA) samples in this $x$ range exhibit magnetic order (spin-glass (SG) like state) and superconductivity, respectively. Therefore, the ground state varied drastically, and the metallic nature was markedly enhanced by annealing~\cite{Asano2019}. Furthermore, from the phase diagram against the actual hole concentration ($p$) estimated by O $K$-edge X-ray absorption spectroscopy, we clarified the increase in the SC transition temperature ($T_{\rm c}$ (onset)) with decreasing $p$ for 0.09 $\leq$ $p$ $\leq$ 0.17~\cite{Asano2020a}. This result is different from the phase diagram of La$_{2-x}$Sr$_x$CuO$_4$ (LSCO), where the optimum $p$ is $\sim$0.16~\cite{Takagi1989}. SC states with higher $T_{\rm c}$ are achieved in the lightly doped region with $p$ $\lesssim$ 0.09, similar to the SC phase diagram of hole-doped $T^{\prime}$-type La$_{1.8-x}$Eu$_{0.2}$(Sr, Ca)$_x$CuO$_4$~\cite{Takamatsu2012, Takamatsu2014, Sunohara2020}. Thus, it is important to clarify the electronic state of LESCO in superconducting $T$*-type $RE_2$CuO$_4$. 

Electronic states in pristine cuprates have been studied intensively through impurity substitution effects on magnetic correlations~\cite{Balatsky2006, Adachi2004, Risdiana2008, Adachi2008, Fujita2008, He2011, Suzuki2012a, Suzuki2012b, Cieplak1993, Suzuki2016, Hiraka2010}. In LESCO, nonmagnetic Zn (Zn$^{2+}$, $S = 0$) substitution effectively enhances the magnetic order in the pristine sample with $p$ $\lesssim$ 0.14, and induces weak static magnetism, even for $p$ $\gtrsim$ 0.14~\cite{Adachi2004, Risdiana2008, Adachi2008}. In addition, the substitution of magnetic impurity Fe (Fe$^{3+}$, $S = 5/2$) strongly stabilizes the magnetic order (SG-like state) over the entire SC phase~\cite{Fujita2008, He2011, Suzuki2012a, Suzuki2012b}. It is worth noting that the magnetic properties stabilized in the underdoped (UD) and overdoped (OD) regions are different. Long-range and short-range magnetic orders are realized in the UD and OD regions, respectively, of Fe-substituted LESCO (Fe-LESCO)~\cite{Cieplak1993, Suzuki2012a}. Furthermore, a complementary study with angle-resolved photoemission and neutron scattering measurements revealed that the spin-density-wave order in OD Fe-substituted LSCO (Fe-LSCO) originates from the Fermi surface inherent in pristine LSCO. In contrast, the localized spins form a one-dimensional stripe-like domain in UD LSCO~\cite{He2011}. From these results, a doping-induced cross-over of the electronic state from the strongly correlated electronic state to the Fermi liquid (FL) state has been discussed~\cite{He2011, Suzuki2016}. 

Following previous works~\cite{Balatsky2006, Adachi2004, Risdiana2008, Adachi2008, Fujita2008, He2011, Suzuki2012a, Suzuki2012b, Cieplak1993, Suzuki2016, Hiraka2010}, we investigated the magnetism in Fe-substituted LESCO (Fe-LESCO) and Zn-substituted LESCO (Zn-LESCO) by $\mu$SR and magnetic susceptibility measurements. 
This paper is organized as follows. We show the experimental procedures and the sample characterizations in Sect. 2 and Sect. 3A, respectively. The results of magnetization and $\mu$SR measurements are presented in Sect. 3B, and the interpretations are described in Sect. 3C. 
Our results elucidate the electronic state in superconducting $T$*-type cuprates. 
To the best of our knowledge, this is the first study on the element substitution effect on magnetism in $T$*-type cuprates. 

\section{Sample preparation and experimental details}

Polycrystals of La$_{1-x/2}$Eu$_{1-x/2}$Sr$_x$Cu$_{1-y}$(Zn, Fe)$_y$O$_4$ with 0.14 $\leq x \leq$ 0.34 and 0.025 $\leq y \leq$ 0.10 were prepared by an ordinary solid-state reaction method from pre-fired powders of La$_2$O$_3$, Eu$_2$O$_3$, SrCO$_3$, Fe$_2$O$_3$, ZnO, and CuO. As in Fe-LSCO, we introduced an additional Sr$^{2+}$ ion to compensate for the reduction of $p$, by substituting Fe$^{3+}$ into the Cu$^{2+}$ site~\cite{Martin_1996}. The value of additionally doped Sr is given by $y$~\cite{Fujita2009a, Suzuki2016}. Since the actual $p$ in LESCO is smaller than $x$, $p$ in La$_{1-(x+y)/2}$Eu$_{1-(x+y)/2}$Sr$_{x}$Cu$_{1-y}$Fe$_y$O$_4$ may shift to the lower doping side with increasing $y$. However, this situation does not defeat the purpose of investigating the electronic state. The mixture of powders was pressed into pellets, and sintered at 1050 $^{\circ}$C in air with intermittent grinding, then annealed in oxygen gas at 40 MPa at 500 $^{\circ}$C for 80 h to obtain OA samples. All samples used in the present study were OA samples. Magnetic susceptibility measurements were carried out under a magnetic field of 500 Oe, using a SC quantum interference device magnetometer (Quantum Design MPMS) to investigate the magnetism in Fe-LESCO and Zn-LESCO. In addition, to visualize the evidence of superconductivity in lightly Fe-substituted LESCO, the shielding and Meissner signals were measured under a field of 10 Oe.
In our oxygen $K$-edge X-ray absorption spectroscopy, the carrier change, that occurs when Eu$^{3+}$ is converted to Eu$^{2+}$, was not observed in LESCO. Therefore, Eu$^{2+}$ does not contribute to the magnetism of LESCO~\cite{Asano2020a}.

We performed zero-field (ZF) $\mu$SR measurements using a pulsed positive muon beam on the S1 spectrometers in the Materials and Life Science Experimental Facility (MLF) in J-PARC, Japan, and a spectrometer (CHRONUS) at Port 4 in the RIKEN-RAL Muon Facility (RAL) in the Rutherford Appleton Laboratory, UK. 
Four pellets with a radius of 15 mm were prepared. Then, considering the beam size with a radius of 10-15 mm, the four pieces were cut into fan shapes to maximize the sample surface area while minimizing the beam hitting anything other than the sample. These four fan-shaped pieces were then arranged in a clover shape on a silver plate.
%Sintered samples were shaped into 90-degree sectors with a diameter of 15mm, and four of these sectors were placed together without any gaps on a silver plate.
All the Zn-LESCO, and a part of the Fe-LESCO with $x = 0,24$ and $y = 0.05$ were measured at MLF. $\mu$SR is a powerful probe for the study of local magnetism, even in samples that are available only in powder form. The asymmetry parameter $A$($t$) ($\mu$SR time spectrum) given by [$F$($t$) - $\alpha$$B$($t$)]/[$F$($t$) + $\alpha$$B$($t$)], where $F$($t$) and $B$($t$) are the total muon events at time $t$ of the forward and backward counters located in the beamline, respectively, was measured. We denote the $A$($t$) normalized by the initial asymmetry $A$($t$ = 0) after correcting the counting efficiencies as $\alpha$. In this paper, we show the normalized time spectra after subtracting the background (BG). Since we observed the spectrum over a long time range, we estimated the constant BG in each Fe-doped sample so that the BGs of the spectra at the characteristic temperatures, that is, high temperature, low temperature, and near spin glass temperature become the same as each other. Then, the temperature-independent BG was subtracted from all the data.

\begin{figure}
\begin{center}
\includegraphics[width=\linewidth]{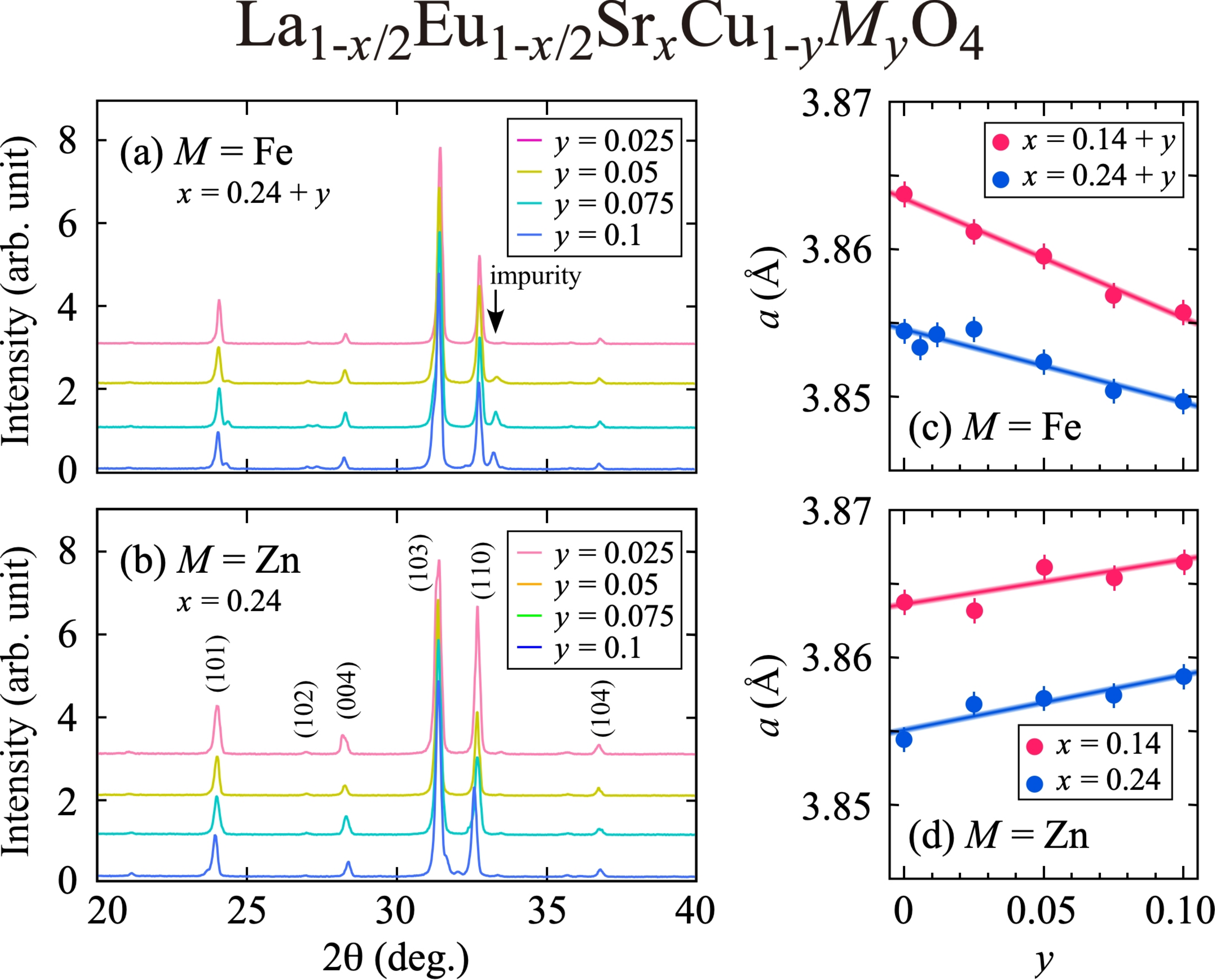}
\end{center}
\caption{\label{lattice-const_v1}
(Color online) XRD pattern and in-plane lattice constant of (a), (c) Fe and (b), (d) Zn-substituted La$_{1-x/2}$Eu$_{1-x/2}$Sr$_x$Cu$_{1-\rm y}$$M_y$O$_4$ with  $x = 0.14 + y$, $x = 0.24 + y$, $x = 0.14$ and $x = 0.24$, respectively. }
\end{figure}
	
\section{Results and discussion}

\subsection{Characterization of samples}

Since there are no studies regarding the element substitution of Cu sites in $T$*-type cuprates, we first examined the structural aspect of the samples. Figures \ref{lattice-const_v1} (a) and (b) show the x-ray powder diffraction patterns of Fe-LESCO with $x = 0.14 + y$ and $0.24 + y$, and Zn-LESCO with $x = 0.14$ and 0.24, respectively, for 0 $\leq$ $y$ $\leq$ 0.10. Fe substitution tends to induce the little amount of the $T^{\prime}$-type phase as the secondary phase, and the sample with $y$ $\geq$ 0.075 contains $\sim$5$\%$ $T^{\prime}$-type phase. All the samples of Zn-LESCO were confirmed to be a single phase of $T$*-type cuprates. We evaluated the in-plane lattice constant ($a$) by Rietveld analysis of the diffraction pattern. The results are plotted in Figs. \ref{lattice-const_v1} (c) and (d) as a function of $y$. The shrinkage (elongation) of the a-axis with increasing $y$ in Fe-LESCO (Zn-LESCO) is caused by replacing Cu$^{2+}$ with Fe$^{3+}$ (Zn$^{2+}$) having a smaller (larger) ionic radius than that of Cu$^{2+}$. Accordingly, the linear variation of the lattice constant suggests that the substitution has been almost achieved, although the impurity phase tends to appear in Fe-LESCO with increasing $y$.

\begin{figure}
\begin{center}
	\includegraphics[width=57mm]{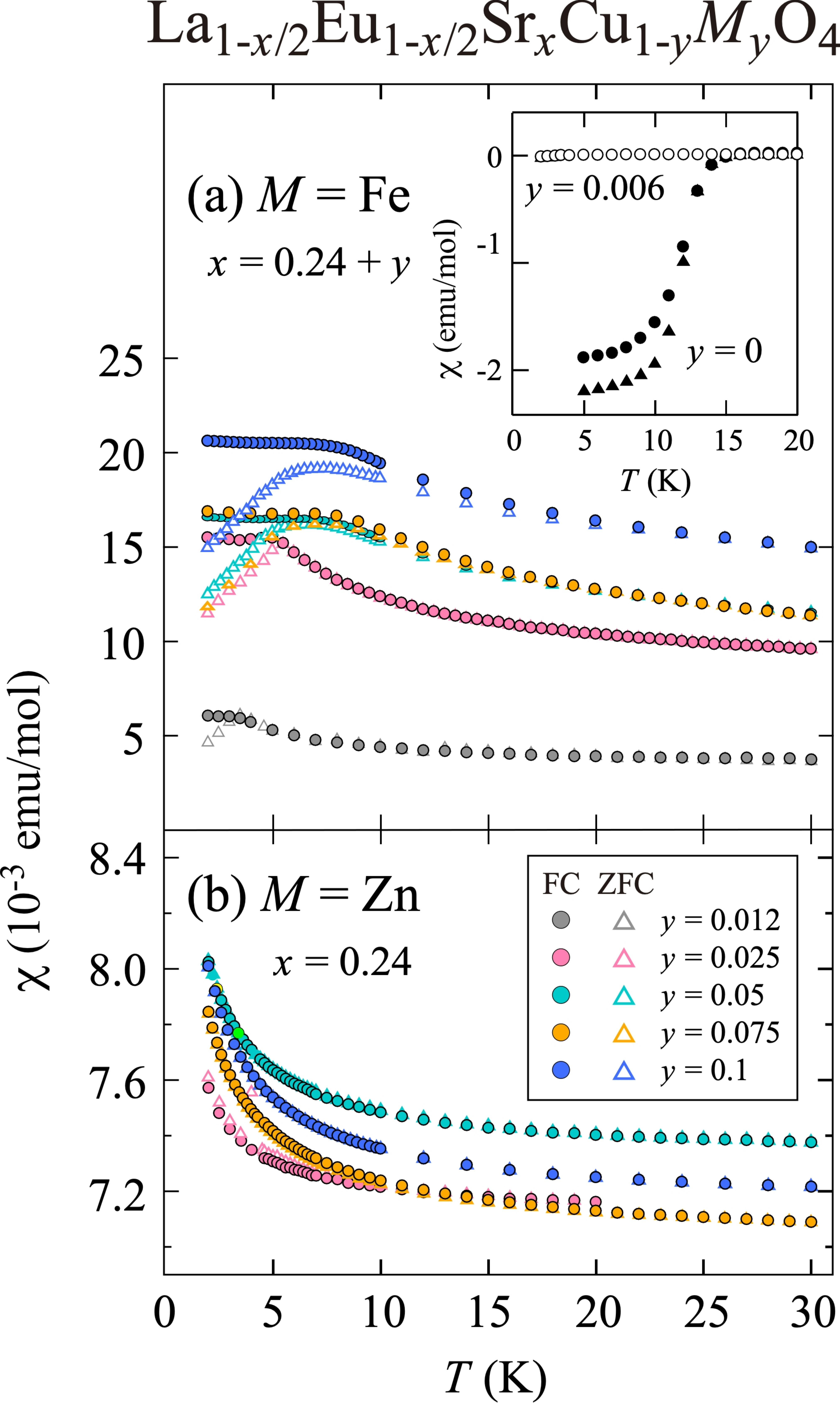}
\end{center}
\caption{\label{susceptibility_v1}
(Color online) (a) Fe- and (b) Zn concentration dependence of magnetic susceptibility measured after zero-field and field-cooling processes for La$_{1-x/2}$Eu$_{1-x/2}$Sr$_x$Cu$_{1-y}$$M_y$O$_4$ with $x=0.24 + y$ for Fe substitution, and $x = 0.24$ for Zn substitution. 
}
\end{figure}

\subsection{Fe and Zn-concentration dependence of magnetism in La$_{1-x/2}$Eu$_{1-x/2}$Sr$_x$Cu$_{1-y}$M$_y$O$_4$}

\subsubsection{Magnetic susceptibility measurements}
First, we examined the effects of Fe- and Zn substitution on the magnetism in LESCO with $x = 0.24$ ($p \sim 0.16$, $T_{\rm c} = 11.9$ K). 
Figs. \ref{susceptibility_v1} (a) and (b) respectively show the temperature dependence of the magnetic susceptibility $\chi$($T$) for Fe-LESCO with $x$ = 0.24 + $y$, and Zn-LESCO with $x$ = 0.24 for $y$ = 0, 0.006, 0.012, 0.025, 0.050, 0.075 and 0.10. In Fe-LESCO, $\chi$($T$) measured after zero-field-cooling (ZFC) and field-cooling (FC) processes ($\chi$($T$)$_{\rm ZFC}$ and $\chi$($T$)$_{\rm FC}$, respectively) exhibits distinct behavior at low temperatures, suggesting the appearance of SG-like magnetic state by Fe-substitution. The $T_{\rm sg}$ where the $\chi$($T$)$_{\rm ZFC}$ exhibits a local maximum slightly increases, and the cusp in $\chi$($T$)$_{\rm ZFC}$ for $y$ = 0.025 turns into a broad peak for $y$ $\geq$ 0.05. The broadening of the cusp suggests the enhancement of magnetic defects for larger $y$. In Zn-LESCO, a paramagnetic behavior exhibiting an increase in $\chi$($T$) upon cooling was observed. Therefore, the magnetic ground state of Fe-LESCO and Zn-LESCO are different. The inset of Fig. \ref{susceptibility_v1} (a) indicates $\chi$($T$)$_{\rm ZFC}$ for Fe-LESCO with, $x$ = 0.24 + $y$ and $y$ = 0 and 0.006. The result for the pristine LESCO ($y$ = 0) is taken from Ref. \onlinecite{Asano2019}. 
Superconductivity is suppressed by the substitution of small amounts of Fe, following the emergence of the SG state.

%%%%Figure 3. muSR time spectra_V1%%%%%
	\begin{figure}[t]
	\begin{center}
	\includegraphics[width=64mm]{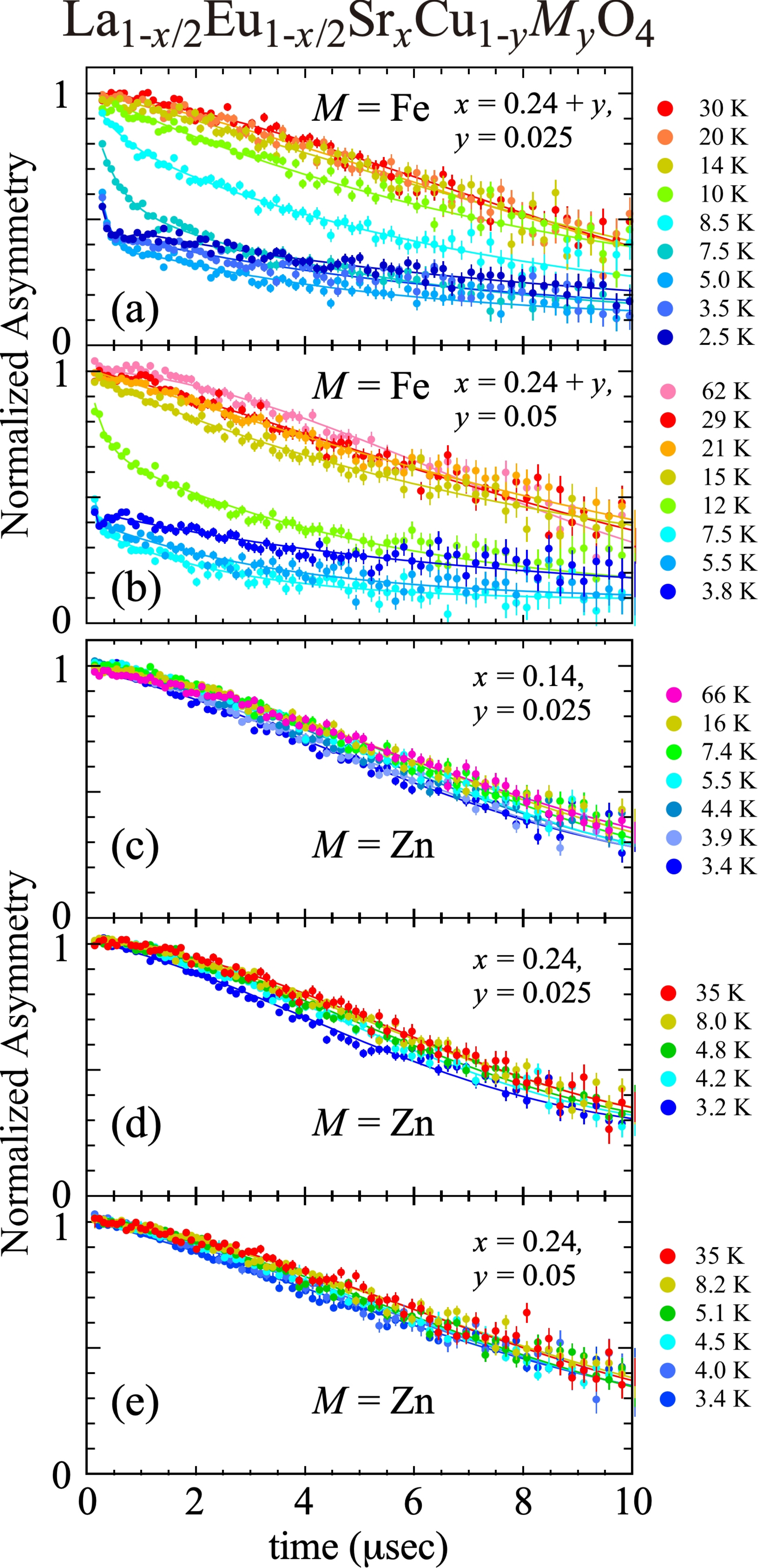}
	\caption{(Color online) $\mu$SR time spectra at zero-field for La$_{1-x/2}$Eu$_{1-x/2}$Sr$_x$Cu$_{1-y}$$M_y$O$_4$ with ($M$, $x$, $y$) = (a) (Fe, 0.265, 0.025), (b) (Fe, 0.29, 0.05), (c) (Zn, 0.24, 0.025), (d) (Zn, 0.24, 0.05), and (e) (Zn, 0.14, 0.025).}
	\label{muSR time spectra_V1}
	\end{center}
	\end{figure}

%%%%Figure 4. spectra_AS_Prm_v1%%%%%
	\begin{figure}[t]
	\begin{center}
	\includegraphics[width=66mm]{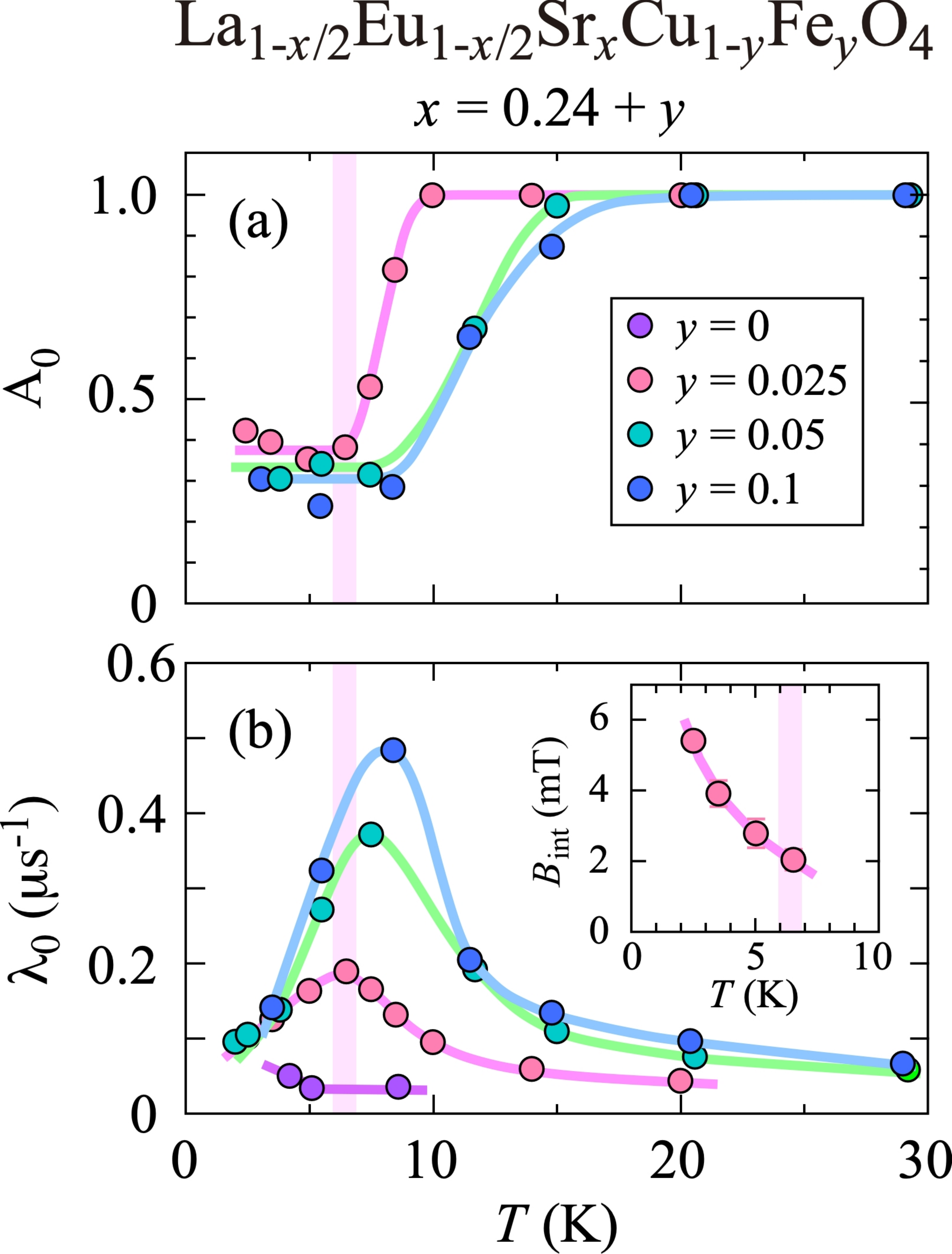}
	\caption{(Color online) The temperature dependence of (a) initial asymmetry $A_0$ and (b) relaxation rate $\lambda _0$ estimated from the $\mu$SR spectra in La$_{1-x/2}$Eu$_{1-x/2}$Sr$_x$Cu$_{1-y}$Fe$_{\rm y}$O$_4$. $\lambda _0$ for $y$ = 0 is obtained from the data in Ref. \onlinecite{Asano2019}. The inset shows the temperature dependence of the internal field $B_{\rm int}$ for $y = 0.025$. }
	\label{spectra_AS_Prm_v1}
	\end{center}
	\end{figure}

\begin{figure}
\begin{center}
\includegraphics[width=65mm]{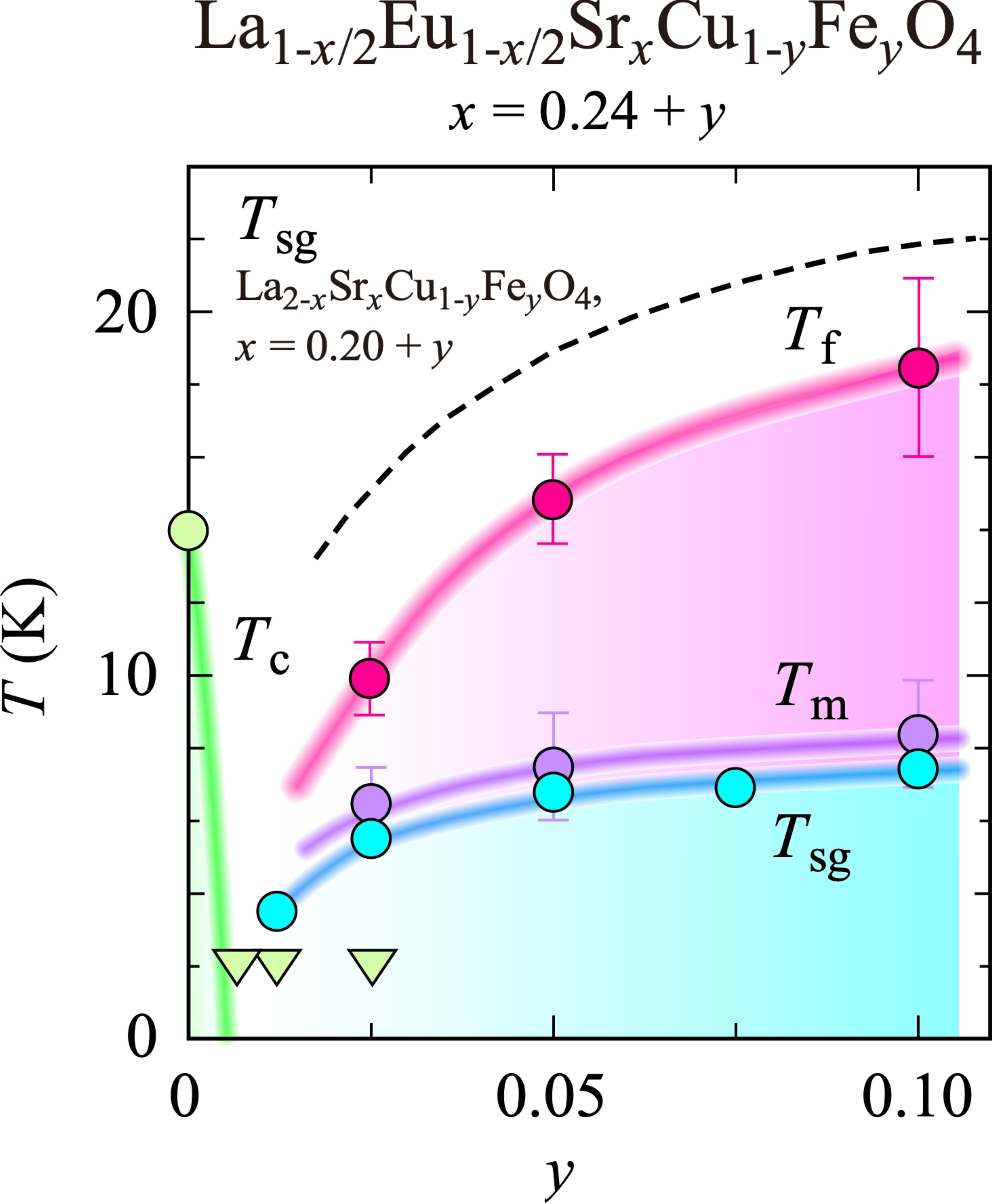}
\end{center}
\caption{\label{phase diagram-Fe_v1}
(Color online) The Fe concentration dependencies of $T_{\rm c}$(green circle and triangles), $T_{\rm m}$(purple circles), $T_{\rm sg}$(light blue circles), and $T_{\rm f}$(pink circles) for La$_{1-x/2}$Eu$_{1-x/2}$Sr$_x$Cu$_{1-y}$Fe$_{\rm y}$O$_4$ with $x$ = 0.24 + $y$. 
$T_{\rm m}$ and $T_{\rm sg}$ are defined as the temperatures at which the temperature dependence of the relaxation rate for the non-magnetic component of the $\mu$SR time spectra and zero-field-cooled magnetic susceptibility respectively show the maximum. $T_{\rm c}$ ($T_{\rm f}$) is evaluated from the results in the inset of Fig. \ref{susceptibility_v1}(a) (Fig. \ref{spectra_AS_Prm_v1}(a)). 
The green triangles means that $T_{\rm c}$ is less than 2 K. 
The dashed line represents $T_{\rm sg}$ for overdoped La$_{2-x}$Sr$_x$Cu$_{1-y}$Fe$_{y}$O$_4$ with $x$ = 0.20 + $y$ taken from Ref. \onlinecite{Suzuki2016}
}
\end{figure}

\subsubsection{$\mu$SR measurements}

Here, we present the results of the $\mu$SR experiments. 
Figure \ref{muSR time spectra_V1} shows the $\mu$SR time spectra for Fe-LESCO with ($x$, $y$) = (a) (0.265, 0.025), (b) (0.29, 0.050), and Zn-LESCO with ($x$, $y$) = (c) (0.24, 0.025) and (d) (0.24, 0.050). In Fe-LESCO, the Gaussian-shaped spectra transformed into an exponential-shaped spectra upon decreasing the temperature, followed by the appearance of the fast depolarization component or oscillation component at low temperatures. A negligible temperature dependence of the Gaussian-shaped depolarization below $\sim$60 K has been reported for the pristine OA LESCO with 0.14 $\leq x \leq$ 0.24~\cite{Asano2019}. This thermal evolution of the spectra in Fe-LESCO means the inducement of the magnetic state by Fe-substitution. 
By contrast, Gaussian-shaped spectra remain at $\sim$3 K in Zn-LESCO, even with $y$ = 0.05, indicating the absence of static magnetism. %However, the spectra slightly depolarizes faster at lower temperatures. 
Thus, as suggested by the susceptibility measurements, the magnetic and non-magnetic ground states are realized by respective Fe and Zn substitution into LESCO with $x = 0.24$. 

To evaluate the magnetic volume fraction ($V_{\rm m}$), internal magnetic field at the muon stopping site ($B_{\rm int}$), and magnetic ordering temperature, we analyzed the $\mu$SR time spectra using the following functions. 
\begin{equation}
\begin{split}
A(t)=A_{0}e^{-\lambda_{0}t}G(\varDelta,t)+A_{1}e^{-\lambda_{1}t}+A_{2}e^{-\lambda_{2}t}\mbox{cos}(\gamma_{\mu}B_{\rm int}t)\label{eq1},
\end{split}
\end{equation}
where
\begin{equation}
G(\varDelta,t)=\frac{1}{3}+\frac{2}{3}(1-\varDelta^{2}t^{2})e^{-\frac{1}{2}\varDelta^{2}t^{2}}. 
\label{eq2}
\end{equation}
$G(\varDelta, t)$ is the Kubo-Toyabe function, and $\varDelta$ represents the distribution of the nuclear dipole field at muon site (full width at half maximum).
The first term of Eq. \ref{eq1} is the non-magnetic component exhibiting slow relaxation due to nuclear dipole fields randomly oriented at the muon site. 
The second and third terms represent the fast depolarization component due to slowly fluctuating electron spins and the magnetically ordered component, respectively. 
Here, $A_0$, $A_1$, and $A_2$ are the initial asymmetry at $t = 0$ corresponding to the fraction of the respective components.
The total initial asymmetry, $A_{\rm tot}$, is defined as $A_{\rm tot}=A_0+A_1+A_2$ and normalized to be 1.
Giving the magnetically ordered powder material, the 1/3 of the spin direction of muons is parallel to the internal field. When the magnetic volume fraction is 100\% and nuclear magnetic fields are negligible compared to that induced by electron spins, this 1/3 component can be described as $A_0 = 1/3$ in Eq. (1).
The second term is used when the rotation component is not observed.
$A_1 + A_2 = \frac{2}{3}$ is a typical constraint for a powder sample and both $A_1$ and $A_2$ can have finite values for systems with inhomogeneous static magnetism.

 Since, we confirmed the temperature-independent $\varDelta$ below 30 K, we fixed $\varDelta$ as the averaged value for our analysis.
$\lambda_0$, $\lambda_1$, and $\lambda_2$ are the relaxation rates. $\gamma_{\mu}$ and $B_{\rm int}$ indicate the gyromagnetic ratio of $\mu^+$ (2$\pi \times$13.55 kHz/G) and the average value of the internal magnetic field at the muon stoping site. 
This suggests that the Cu spins are partially ordered around the Fe ions, while most of the static magnetism corresponds to a spin glass state.
Here, we assume a single muon stopping site, since the above equations can well reproduce the observed spectra. 

Figures \ref{spectra_AS_Prm_v1}(a) and (b) show the temperature dependence of $A_0$ and {$\lambda_0$} respectively for Fe-LESCO, with $x$ = 0.24 + $y$ and $y$ = 0, 0.025, 0.050, and 0.10. 
The inset of Fig. \ref{spectra_AS_Prm_v1} (b) shows $B_{\rm int}$ for $y$ = 0.025. The results for pristine LESCO with $x = 0.24$ were adopted from Ref. \onlinecite{Asano2019}. Upon cooling, $A_0$ for all the Fe-doped samples starts to decrease at 10 -- 15 K. With further cooling, $A_0$ is reduced to $\sim$1/3, corresponding to the value for the full magnetic volume fraction. Since the amount of doped Fe is 10\% at most, the bulk magnetic order is not caused solely by Fe$^{3+}$ spins, which are diluted on the CuO$_2$ planes. 

The temperature for the maximum $\lambda_0$ ($T_{\rm m}$) is comparable to the temperature at which $A_0$ reaches an almost constant value, and the weak oscillation component appears upon cooling. Therefore, the freezing of the spins gradually occurs at lower temperatures. The increase in $B_{\rm int}$ below $T_{\rm m}$ indicates an increase in the moment size of the Cu spins in the ordered phase, although $V_{\rm m}$ is almost 100 $\%$. However, the significant damping of the spectra suggests a rather short-range magnetic order with a broad distribution of the internal magnetic fields at the muon stopping sites, even at low temperatures. The same trend was obtained for the $y$ = 0.050 and 0.10 samples. 
If the $A_2$ term is ignored, the behavior less than 2 $\mu$s cannot be reproduced in the Fe sample at the lowest tempereature. 

\subsubsection{Magnetic state in LESCO with $x$ = 0.24 ($p$ $\sim$ 0.16)}
	
Fig. \ref{phase diagram-Fe_v1} summarizes the $y$-dependence of $T_{\rm c}$, $T_{\rm m}$ and $T_{\rm sg}$ for La$_{1-x/2}$Eu$_{1-x/2}$Sr$_x$Cu$_{1-y}$Fe$_y$O$_4$ with $x$ = 0.24 + $y$. $T_{\rm sg}$ for $T$-type Fe-LSCO with $x$ = 0.20 + $y$ is shown by a dashed line as a reference~\cite{Suzuki2016}. 
The green triangles means that $T_{\rm c}$ is less than 2 K. The onset temperature for the development of $V_{\rm m}$ upon cooling ($T_{\rm f}$) estimated from Fig. \ref{spectra_AS_Prm_v1}(a) is also plotted. $T_{\rm m}$ is slightly higher than $T_{\rm sg}$ at each concentration, and both $T_{\rm m}$ and $T_{\rm sg}$ increase weakly with increasing $y$. $T_{\rm f}$ is approximately double of $T_{\rm m}$. This difference in the characteristic temperatures can be attributed to the experimental probes with different time windows to observe SG transition~\cite{Wakimoto2000, Enoki2011, Fujita2008, He2011, Suzuki2016}. 

Here, we discuss the origin of the magnetic properties and electron states of LESCO by comparing them with $T$-type copper oxide. 
The results for Fe-LESCO are similar to those observed for OD Fe-LSCO in the following aspects: the emergence of bulk magnetic order by a partial Fe substitution, fast depolarization in the low temperature $\mu$SR spectra, 
and increase in $T_{\rm m}$ with increasing $y$. These characteristics are consistent with the presence of polarized itinerant spins in the SG state due to the Ruderman-Kittel-Kasuya-Yosida (RKKY) interaction in a metallic state, as indicated by the OD Fe-LSCO~\cite{Suzuki2012a, Suzuki2016} and OD Fe-substituted Bi$_{1.75}$Pb$_{0.35}$Sr$_{1.90}$CuO$_{4+\delta}$~\cite{Hiraka2010, Wakimoto2010}. 
This implies that the randomly distributed magnetic moments on the CuO$_2$ planes cause fast depolarization in the $\mu$SR time spectra, and $T_{\rm m}$ tends to increase with an increase in the number of Fe ions, whose moments interact via itinerant carriers. Contrary to this, the strong damping of the $\mu$SR time spectra is different from the observation of a clear oscillation component for UD Fe-LSCO~\cite{Suzuki2016, Suzuki2012b}, in which long-range magnetic and charge stripe orders coexist\cite{Fujita2009a, Fujita2009b}. 

As discussed in Refs. \onlinecite{He2011} and \onlinecite{Suzuki2016}, better nesting conditions in OD samples enhance the spin polarization of the doped holes and stability of the magnetic order due to the RKKY interaction. In this context, the lower $T_{\rm m}$ in $T$*-type Fe-LESCO than that in $T$-type Fe-LSCO with comparable $p$ and $y$~\cite{Suzuki2012a, Suzuki2012b, Suzuki2016} suggests weaker RKKY interaction and nesting conditions in Fe-LESCO. The negligible Zn substitution effect on the static magnetism is similar to the results for OD Zn-LESCO, but different from the enhancement of the magnetic order by Zn-substitution seen in UD LSCO~\cite{Adachi2004, Risdiana2008, Adachi2008}. Therefore, as is the case for the OD region of $T$-type LSCO, the itinerant spin nature is dominant in the present Fe-LESCO, and FL is most likely the ground state in the pristine $T$*-type LESCO with $p$ $\sim$ 0.16. 

\subsection{Sr-concentration dependence of the magnetic state in La$_{1-x/2}$Eu$_{1-x/2}$Sr$_x$Cu$_{1-y}$(Fe, Zn)$_y$O$_4$}

\subsubsection{Magnetic susceptibility and $\mu$SR measurements}

\begin{figure}
\begin{center}
\includegraphics[width=67mm]{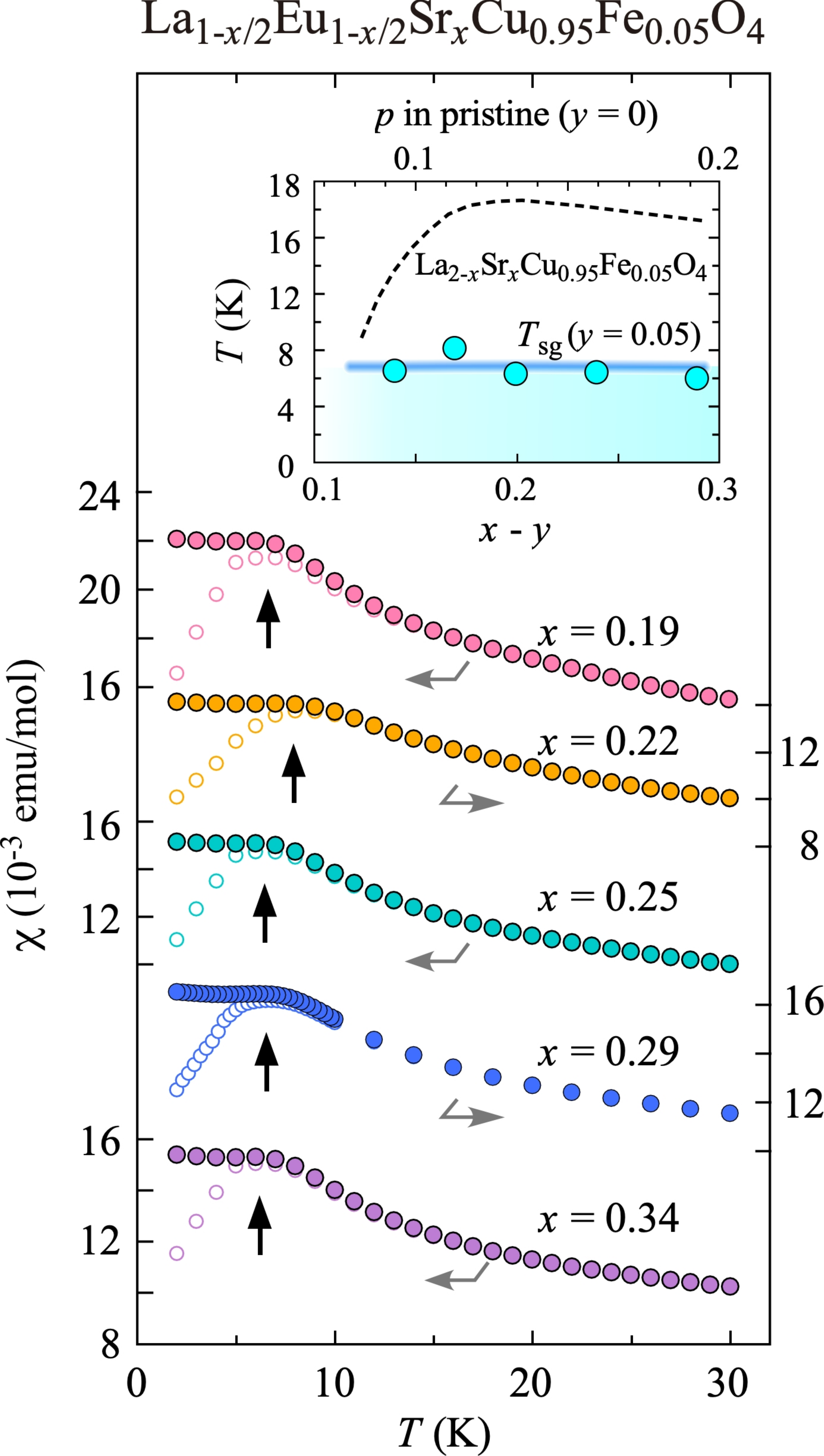}
\end{center}
\caption{\label{susceptibility_Fe-const-y_v1}
(Color online) Temperature dependence of the magnetic susceptibility ($\chi$($T$)) measured after FC and ZFC processes in a magnetic field of 500 Oe for La$_{1-x/2}$Eu$_{1-x/2}$Sr$_x$Cu$_{1-y}$Fe$_y$O$_4$ (Fe-LESCO) with $x = 0.19$. 0.34 and $y = 0.05$. $\chi$($T$) are shifted for clear visualization. Arrows represent $T_{\rm sg}$, where ZFC $\chi$($T$) is the maximum. The dashed line in the inset represents $T_{\rm sg}$ for La$_{2-x}$Sr$_x$Cu$_{0.95}$Fe$_{0.05}$O$_4$ taken from Ref. \onlinecite{Suzuki2016}}
\end{figure}

Next, we examined the $x$-dependence of the magnetization for 5 $\%$ Fe-substituted LESCO with 0.19 $\leq$ $x$ $\leq$ 0.34 to investigate the extent to which the plausible FL state extends against $x$. As seen in Fig. \ref{susceptibility_Fe-const-y_v1}, all the samples exhibit hysteresis in $\chi$($T$), indicating the inducement of the SG state by Fe substitution in a wide $x$ range. 
Thus, $T_{\rm sg}$ is almost constant against $x$ in the range between 0.19 and 0.34, as seen in the inset of Fig. \ref{susceptibility_Fe-const-y_v1}. In contrast, in 5$\%$ Fe-LSCO (dashed line), $T_{\rm sg}$ increases rapidly from $\sim$10 K for $x = 0.18$ to $\sim$18 K for $x = 0.24$, and saturates above $x = 0.24$~\cite{Suzuki2016}. The rapid change in $T_{\rm sg}$ seems to be associated with the cross-over of the ground state from the strongly correlated electronic state to the FL state~\cite{He2011, Suzuki2016}. 

The constant value of $T_{\rm sg}$ in LESCO similar to the case of Fe-LSCO suggests the existence of the same electronic state, namely, the FL state in a broad $x$ range. We note that evidence of successive magnetic transitions from the SG state followed by the spin-stripe ordered state upon cooling was reported for optimally doped and OD Fe-LESCO~\cite{Suzuki2012a, Suzuki2016}. That is, the $\chi$($T$)$_{\rm ZFC}$ and the temperature dependence of $A_0$ determined by $\mu$SR exhibits two local maxima and two peaks, respectively ~\cite{Suzuki2012a, Suzuki2016}. However, in the present Fe-LESCO, no such behavior was observed for all the samples down to $\sim$2 K, which is consistent with the absence of a stripe ordered state. 

To further confirm the absence of a strongly correlated electronic state in $T$*-type LESCO with lower $p$, we measured the 2.5$\%$ Zn-substitution effect on the magnetism in LESCO with $x = 0.14$ ($p \sim 0.09$, $T_{\rm c} = 24.5$ K) by additional $\mu$SR experiments. In LSCO, a small amount of Zn substitution samples showed the stripe order in $p$ $<$ 0.14, but induces only weakly static magnetism in the FL state of the OD region~\cite{Adachi2004, Risdiana2008, Adachi2008}. Therefore, if a strongly correlated electronic state is present in the lower $p$ region, the enhancement of static magnetism could be observed by Zn substitution. Fig. \ref{muSR time spectra_V1} (e) shows the thermal evolution of the observed $\mu$SR time spectra. The results are approximately identical to those of the temperature dependence for $x = 0.24$ ($p \sim 0.16$) and $y = 0.025$. The Gaussian-shaped spectra did not change with temperature, indicating the absence of static magnetism. Weak thermal evolution of the $\mu$SR time spectra with the remaining Gaussian-shaped depolarization was also observed in Zn-LESCO with $x = 0.14$ and $y = 0.05$ and 0.10 (not shown here).

\subsubsection{Relevance to $T$- and $T^{\prime}$-type cuprates}

In the present study, we obtained no evidence of cross-over of the electronic state against doping and successive magnetic order with decreasing temperature. Considering the above observations, we conclude that the SG-like state is most likely originated by RKKY interaction, and the FL state is realized in the present $T$*-type LESCO with $x$ $\geq$ 0.14. Even when a stripe ordered phase exists, it would emerge in a more lightly doped region of the LESCO. Since the actual $p$ value of 0.09 is reported for pristine LESCO with $x = 0.14$~\cite{Asano2020a}, it can be concluded that the FL state is formed in the lower $p$ region, contrary to that in $T$-type LESCO. 
The formation of the FL state with lower $p$ can be attributed to the smaller size of the charge-transfer gap ($\Delta_{\rm CT}$) in the parent $T$*-type $RE$CuO$_4$. From the optical conductivity measurements on various types of Cu-O single-layer networks, Tokura and co-workers clarified the smaller $\Delta_{\rm CT}$ for $RE_2$CuO$_4$ with lower oxygen coordination~\cite{Tokura1990}. Carrier doping into the system with smaller $\Delta_{\rm CT}$ could induce the metallic state at lower $p$, because the charge transfer gap collapses easily with filling the in-gap state. Such a trend is reported for $T^{\prime}$-type $RE_2$CuO$_4$\cite{Naito2002, Fujita2003, Krockenberger2008, Krockenberger2014} via Ce-substitution. The onset electron concentration for the emergence of superconductivity in $RE_2$Ce$_x$CuO$_4$ is lower in the system with a smaller $\Delta_{\rm CT}$ in the parent compound. It is worth mentioning that Ce-free $T^{\prime}$-type La$_{1.8}$Eu$_{0.2}$CuO$_4$, which has a large in-plane lattice constant (4.00 $\AA$), exhibits superconductivity after adequate oxygen reduction annealing~\cite{Takamatsu2012}. From the negative linear relation between $\Delta_{\rm CT}$ and the in-plane lattice constant, La$_{1.8}$Eu$_{0.2}$CuO$_4$ is expected to have the smallest $\Delta_{\rm CT}$ among the $T^{\prime}$-type $RE_2$CuO$_4$~\cite{Arima1991, Asano2020b}. Furthermore, the increase in $T_{\rm c}$ with lowering $p$ in $T$*-type LESCO is similar to the $p$-dependence of $T_{\rm c}$ in the hole-doped $T^{\prime}$-type La$_{1.8-x}$Eu$_{0.2}$(Sr, Ca)$_x$CuO$_4$~\cite{Takamatsu2012, Takamatsu2014, Sunohara2020}. This similarity suggests that the doping evolution of the electronic state in LESCO with five oxygen coordination is comparable to that in La$_{1.8-x}$Eu$_{0.2}$(Sr, Ca)$_x$CuO$_4$ with four coordination, than that in LSCO with six coordination. 
Of all the possibilities, the above discussion is the most likely candidate.
Thus, our results suggest the importance of research on the lightly doped $T$*-type cuprate for a unified understanding of the ground state in $RE_2$CuO$_4$, and the mechanism of superconductivity from the viewpoint of Cu coordination. 

\section{Summary}

We have investigated the magnetism in Fe- and Zn-substituted $T$*-type LESCO by $\mu$SR and magnetic susceptibility measurements for the first time to clarify the ground state in pristine LESCO. 
SG-like field-dependent behavior was observed in the temperature dependence of the magnetic susceptibility $\chi$($T$) in Fe-LESCO. Using $\mu$SR measurements, we confirmed the existence of bulk magnetic order in all Fe-LESCO samples, and a slight increase in the magnetic ordering temperature with Fe substitution. These results are quantitatively similar to the observation in OD $T$-type Fe-LSCO, for which the RKKY interaction was discussed as the origin of the magnetic order. Contrary to this, in Zn-LESCO, the paramagnetic behavior in $\chi$($T$) with no magnetic order was observed. All of these element substitution effects are consistent with those found in the OD LSCO~\cite{Suzuki2012a, Suzuki2016}, suggesting that the FL ground state is realized in pristine OD LESCO from a lower $p$ upon doping, compared to that in $T$-type LSCO. The existence of the FL state in the lower $p$ region is possibly associated with the smaller size of the charge transfer gap in the $T$*-type LESCO than in the $T$-type LSCO, owing to the lower number of oxygen coordinations. 
Our result suggests that the local crystal structure around the CuO$_2$ plane has a strong influence on the size of the charge transfer gap and plays a vital role in the mechanism of superconductivity in $T$*-type LESCO. \\

\section*{Acknowledgements}

We thank T. Adachi, Y. Koike, and Y. Miyazaki for fruitful discussions and appreciate T. Noji and T. Kawamata for their help in sample preparation. 
The $\mu$SR experiments at the Materials and Life Science Experimental Facility of J-PARC (Proposal Nos. 2015MP001, 2018B0324, and 2019B0290) and at the RIKEN-RAL Muon Facility in the Rutherford Appleton Laboratory (Proposal No. RB1970107) were performed under user programs. 
We thank the J-PARC and the RIKEN-RAL staff for their technical support during the experiments. This work was partially supported by the IMSS Multiprobe Research Project, and M.F. and T.T. are supported by Grant-in-Aid for Scientific Research (A) (16H02125) and for Young Scientists(Start-up) (19K23417), respectively.

\end{document}